\documentclass[prb,twocolumn,
			   showpacs,floatfix,
			   superscriptaddress]{revtex4}

\usepackage{amssymb}
\usepackage{graphicx}
\usepackage{dcolumn}
\usepackage{bm}
\usepackage{latexsym}
\usepackage{amsmath}

\begin{document}

\title{Electron pairing in one-dimensional quasicrystals}

\author{Y. \surname{Arredondo}}
\email{yesenia@iim.unam.mx}
\affiliation{Instituto de Investigaciones en Materiales,
             Universidad Nacional Aut\'onoma de M\'exico, 
             Apartado Postal 70-360, 04510 M\'exico D.F., 
			 M\'exico.}

\author{O. \surname{Navarro}}
\affiliation{Instituto de Investigaciones en Materiales,
             Universidad Nacional Aut\'onoma de M\'exico, 
             Apartado Postal 70-360, 04510 M\'exico D.F., 
			 M\'exico.}
\affiliation{Instituto de Investigaciones Metal\'urgicas,
             Universidad Michoacana de San Nicol\'as Hidalgo, 
             Edificio ``U'' Ciudad Universitaria, 58000 Morelia 
			 Michoac\'an, M\'exico.}			 

\begin{abstract}
Electron pairing in one-dimensional binary Hubbard chains is 
studied for different values of the band-filling using the 
Density Matrix Renormalization Group method. The systems consist 
of linear arrays of sites with two types of on-site correlations 
defined by two potentials: $U_A$ being attractive ($< 0$), 
and $U_B$ repulsive ($> 0$). The atomic levels of the system 
are modulated with periodic and quasiperiodic ordering, in the 
latter case following the Fibonacci sequence. We analyze the effect 
of such modulations and calculate the electron pairing 
phase diagram as a function of the band-filling. It is observed
that there is a critical value of the band-filling where the behavior
of the periodic and the Fibonacci binary Hubbard chains is reversed.

\end{abstract}
\pacs{71.10.Fd, 71.10.Li, 71.23.Ft}

\maketitle

An impression that low-dimensional systems are an ideal
and purely serve as a mathematical example is vanishing
from stage. The several nanoscopic
low-dimensional devices that have been obtained in the
laboratories such as carbon nanotubes, semiconducting
quantum wires and quantum dots mean a breakthrough in the 
engineering of materials with novel physical properties;
with desired properties such as superconductivity, and with 
promising applications, for example, in the area of spintronics. 
Such systems are inhomogeneous in their structure and
intrinsically strongly correlated due to the reduction of the
available phase space.
To this collection of materials we can add the one-dimensional 
(1D) quasicrystals (QCs). One of the best known 1D quasicrystals is based on the
Fibonacci sequence which started drawing interest after the
papers by Kohmoto et al. \cite{kk83} and Ostlund et al. \cite{so83}. The
spectral properties of the Fibonacci chain are exotic; the
single-particle eigenstates are neither extended nor localized but critical
and the spectrum corresponds to that of a Cantor set \cite{kk83,so83,ko84,wf93}. 
A Fibonacci sequence
consists of two elements $A$ and $B$ and the entire sequence is generated by
successive application of the substitution rule. The first few generations
are $G_{0}=B,$ $G_{1}=A,$ $G_{2}=AB,$ $G_{3}=ABA,$ $G_{4}=ABAAB,$ $...,$ $%
G_{i}=G_{i-1}G_{i-2}$ for $i\geq 2,$ where the letter $G_{i}$ indicates the
$i$th generation. In a Fibonacci chain, the elements $A$ and $B$ from the
Fibonacci sequence may denote two different atoms (diagonal model) or two
different bonds separating identical atoms (off-diagonal model). 
In this work, we will study the diagonal model, where the site energy 
takes two values $\varepsilon_{A}$ and $\varepsilon_{B}$ 
associated to atoms $A$ and $B$, respectively. 
In the corresponding diagonal model the number of sites with 
energy $\varepsilon_{A}$ is $N_{A}(n)$ and the number of sites 
with energy $\varepsilon_{B}$ is $N_{B}(n)$. 
The total number of sites in a generation $n$ is represented by 
$N(n)$, $N(0)=N(1)=1$. These numbers are related by
\begin{eqnarray*}
N(n) &=&N(n-1)+N(n-2),  \nonumber \\
N_{A}(n) &=&N(n-1),  \label{fib-seq} \\
N_{B}(n) &=&N(n-2).  \nonumber
\end{eqnarray*}
In the quasiperiodic limit $(n\rightarrow \infty )$ the ratio 
$N_{A}(n)/N_{B}(n)$ converges towards the golden mean $\sigma =(\sqrt{5}+1)/2$.
Motivated by the work of Alexandrov et al. \cite{Alexandrov}
on the $s-$wave electron pairing in a one-dimensional binary Hubbard system 
with two electrons and a total momentum $\mathbf{K}=0$, 
we investigate the electron pairing for different values 
of the band-filling in a Fibonacci lattice. 
In their paper, Alexandrov et al. considered the possibility
of a binary system due to the fact that the $CuO$ chains 
play a key role in the high-temperature superconductivity. 
Thus, they considered a Hubbard model with repulsion on copper 
and attraction on oxygen atoms. And for simplicity, 
the energy of the atomic levels was set the same for all sites. 
We further want to study the interplay of the binary Hubbard systems 
under the influence of a modulation of the atomic energy levels, which we
model by setting their values along the chain distributed according to the 
Fibonacci sequence. The results are compared to those of the homogeneous and 
diatomic chains. The binary Hubbard system proposed by Alexandrov with only 
two electrons has been solved by means of the Bethe Ansatz method. In our case,
we use the Density Matrix Renormalization Group method
(DMRG) \cite{92White, 92BWhite, 98Peschel, 08Arredondo} 
which is an efficient method
for investigating low-energy properties of many-body and strongly 
correlated systems such as those briefly described above.

A one-dimensional binary Hubbard chain consists of non-equivalent 
sites distinguished through two on-site potentials: 
$U_A$ being attractive ($< 0$), 
and $U_B$ repulsive ($> 0$) arranged in alternating order.
The one-dimensional Hubbard model is one of the few
examples of an exactly solvable model using 
Bethe Ansatz \cite{LiebWu}, where the $N$-particle
wave function is constructed using plane-wave exponents
with coefficients obtained from a two-particle S-matrix.
The binary Hubbard system in Ref.~\onlinecite{Alexandrov} was formed
by a collection of unit cells consisting 
of two sites: $A$ and $B$ (see Figure 1a).
For the $i$th unit cell the creation (annihilation) operators
with spin $\sigma$ ($=\uparrow,\downarrow$) are $a_{i,\sigma}^{\dagger}$
($a_{i,\sigma}$) and $b_{i,\sigma}^{\dagger}$ ($b_{i,\sigma}$) at sites
$A$ and $B$, respectively. The following Hamiltonian was then proposed:
\begin{eqnarray*}
  H = -t \sum_{i,\sigma}\Big\{ 
  	  a_{i,\sigma}^{\dagger}b_{i-1,\sigma} +
  	  b_{i,\sigma}^{\dagger}a_{i+1,\sigma} +	
  	  a_{i,\sigma}^{\dagger}b_{i,\sigma} +
  	  b_{i,\sigma}^{\dagger}a_{i,\sigma}
  	  \Big\}  \\
	+ \sum_{i}\Big\{
      U_An^{a}_{i\uparrow}n^{a}_{i\downarrow} +
      U_Bn^{b}_{i\uparrow}n^{b}_{i\downarrow} 
  	  \Big\},
  \label{Alexandrov}
\end{eqnarray*}
with $\varepsilon_A-\varepsilon_B=0$ and two electrons in the chain. 
$n^{a}_{i,\sigma}$ and $n^{b}_{i,\sigma}$ are the
the electron number operators on sites $A$ and $B$, respectively,
with $U_A$ the attractive potential on site $A$ and $U_B$ the
repulsive potential on site $B$.
From the Bethe Ansatz solution it was then concluded
that, given a value $U_B$ for the repulsive interaction,
there is a critical value of $U_A$ above which
pairing of the electrons was always taking place 
in the system. The value of the attractive interaction
is given by:
\begin{equation}
  \vert U_A \vert \ge \frac{2 U_B}{U_B + 2}.
  \label{Bethe}
\end{equation}

\noindent In Ref.~\onlinecite{Alexandrov} it was also left 
open what would happen for
systems with more than two electrons. One possibility mentioned there
was that doping the systems would render in less paired-states,
which would agree with the fact that for high-temperature 
superconductors there is a saturation in their critical
temperature when doping these materials. 

We want now to extend the investigation of electron pairing to
binary Hubbard systems with more than two particles and to lattices with 
different topology like the Fibonacci one. We propose the
following Hamiltonian:
\begin{equation}
  H = \sum_{i\sigma}\varepsilon_i n_{i,\sigma}
  -t \sum_{i,\sigma}c_{i,\sigma}^{\dagger}c_{i+1,\sigma} + H. c. +
  \sum_{i}U_in_{i\uparrow}n_{i\downarrow}, 
  \label{Hubbard}
\end{equation}
where $c_{i,\sigma}^{\dagger}$ ($c_{i,\sigma}$) is the creation 
(annihilation) operator with spin $\sigma$ ($=\uparrow,\downarrow$) 
at site $i$ and $n_{i\sigma}=c_{i,\sigma}^{\dagger}c_{i,\sigma}$ 
is the electron number operator. $\varepsilon_i$ is the atomic energy
level and $t=1$ is the nearest neighbor 
hopping matrix, which we choose to set the energy scale. 
$U_i$ is the on-site interaction and will take the value
$U_A$ for $i$ odd  and $U_B$ for $i$ even.
The Hamiltonian in Eq.~(\ref{Hubbard}) incorporates the
different systems we want to study. We want to compare three different
atomic level scenarios: $i)$ The homogeneous case, which corresponds
to the system studied in Ref.~\onlinecite{Alexandrov}, 
and in which $\varepsilon_i=\varepsilon$, 
i.e., the atomic level is set the same for all sites and we set to zero. 
$ii)$ The diatomic case, where there are
two different values intercalated,  $\varepsilon_A$ and  $\varepsilon_B$ 
( $\varepsilon_A < \varepsilon_B$). In this case, half of the
sites have an atomic level of $\varepsilon_A$ and half of them have
the value $\varepsilon_B$. $iii)$ The last system is the Fibonacci case, where 
we assign each site an atomic energy value following the Fibonacci sequence
starting with $\varepsilon_A$ and thus we will have more sites 
with $\varepsilon_A$ than with $\varepsilon_B$. The exact number 
depends on the generation of the Fibonacci set to be used. Figures
1a, 1b, and 1c show the atomic level structures just described.
\begin{figure}[ht]
	\includegraphics[scale=0.2]{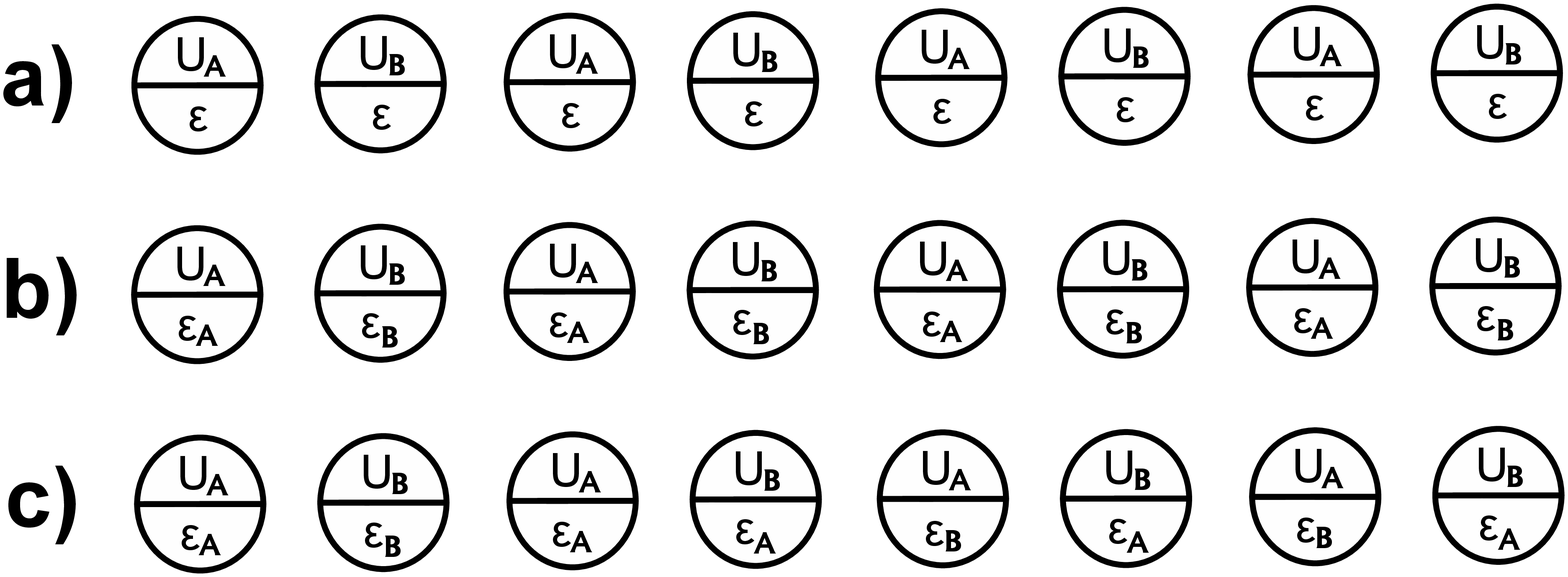}
	\caption{Structure of one-dimensional binary Hubbard Systems.
			 a) The homogeneous binary Hubbard chain, with 
			 $\varepsilon_A = \varepsilon_B=\varepsilon$. 
			 b) The diatomic binary Hubbard chain, with 
			 $\varepsilon_A \neq \varepsilon_B$. 
			 c) The Fibonacci binary Hubbard chain, with 
			 $\varepsilon_A \neq \varepsilon_B$. 
			 Up to the $5$th Fibonacci generation is shown
			 in this diagram and
			 $\varepsilon_A < \varepsilon_B$ for b) and c).}
    \label{fig:Systems}
\end{figure}
It is expected that the quantum confinement originated in the
diatomic and Fibonacci chains due to the fact that 
$\varepsilon_A < \varepsilon_B$ will, in general, 
enhance localization and therefore favor the formation 
of local electron pairs.
On the other hand, the total effect
of the different local potentials and confinement might be cumbersome.
Furthermore, at half-filling Umklapp scattering 
plays a key role in the
electronic properties of 1D systems.
As already mentioned, the interesting
materials and systems have an inhomogeneous
structure, which on the atomic level induces
potentials that modify their properties.
In order to investigate such effects we need to
consider the strong correlations in many-body
quantum systems. Lacking an analytical solution,
we make our attempts using the density matrix 
renormalization group method (DMRG).
This method has its origin in the numerical renormalization
group formulated by Wilson \cite{Wilson} and
allows for accurate calculations of
ground state properties in low-dimensional quantum lattice systems
for which the basis of the Hilbert space grows exponentially
with the number of particles and cannot be handled using
exact diagonalization.
The DMRG is a variational, real-space method that selects 
in a systematic way a sector of the Hilbert 
space that best represents the ground state of a system, which is
done by selecting only the $m$ most probable states calculated from
the density matrix of the system. 
The numerical error caused by truncation of the original basis can be 
calculated directly as the total weight of the states that were 
discarded.  For our systems, with $L=144$ sites, 
we kept $m=256$ density-matrix states, 
resulting in a maximum truncation error of the order of $10^{-6}$.
The length of the systems corresponds to the $12$th Fibonacci 
generation, being $\varepsilon_B$ the zeroth generation 
and $\varepsilon_A$ the first one.
The systems above described were investigated
for different values of the band-filling under open boundary 
conditions, which favors convergence in 
the DMRG method. Finite-size effects such as particle density
oscillations and charge accumulation close to the edges of the systems
are present specially in the homogeneous system away from
half-filling. The diatomic and 
Fibonacci chains behave according to the energy considerations
of their arrangements even at the edges of the systems.

\begin{figure}[ht]
  	\includegraphics[angle=270,scale=0.33]{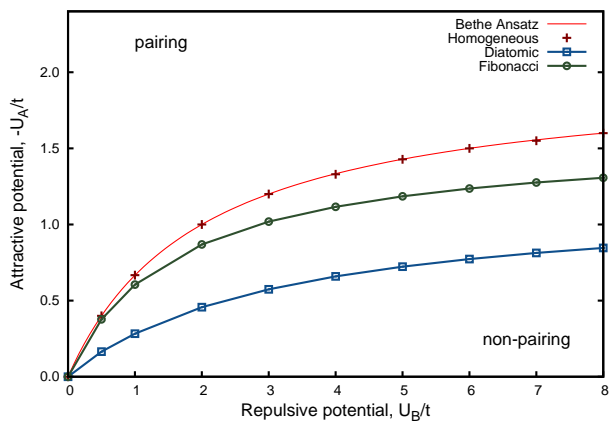}
    \caption{Ground state pairing phase diagram for binary 
    		 Hub\-bard systems with 2 electrons.}
    \label{fig:phase2}
\end{figure}

\begin{figure}[ht]
	\includegraphics[angle=270,scale=0.33]{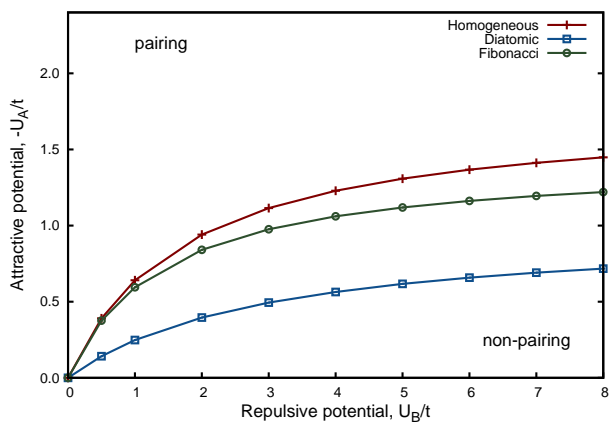}
    \caption{Ground state pairing phase diagram for binary 
    		 Hub\-bard systems at quarter-filling.}
    \label{fig:phase72}
\end{figure}

\begin{figure}[ht]
	\includegraphics[angle=270,scale=0.33]{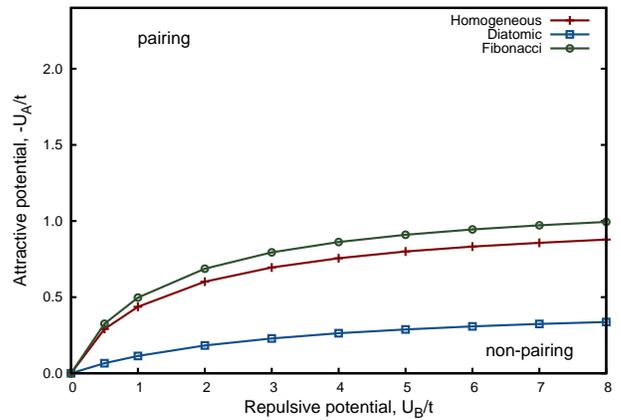}
    \caption{Ground state pairing phase diagram for binary 
    		 Hub\-bard systems at half-filling.}
    \label{fig:phase144}
\end{figure}

To obtain the pairing phase diagram we calculate the binding energy 
$\Delta$ as the difference in the ground state energy $E_{GS}$ 
when the on-site potentials are all off and after they were switched on:
\begin{eqnarray}
  \Delta\big(f(\varepsilon)\big) = 
  		   E_{GS}\Big(U_A=U_B=0; f(\varepsilon)\Big) \nonumber \\
         - E_{GS}\Big(U_A \neq U_B\neq 0; f(\varepsilon)\Big),
\end{eqnarray}
where $f(\varepsilon)$ refers to the homogeneous, diatomic or Fibonacci
arrangement of the atomic levels $\varepsilon$.
A positive $\Delta$ means there is local pair formation in the system.
In Figures \ref{fig:phase2}$-$\ref{fig:phase144} we show the ground state
phase diagram for binary Hubbard chains with homogeneous, diatomic and
Fibonacci ordering of the site energies $\varepsilon$. 
In all cases, the area above (and including) each line corresponds 
to paired states. The area below the line corresponds to non-paired states. 

\begin{figure}[ht]
	\includegraphics[angle=270,scale=0.33]{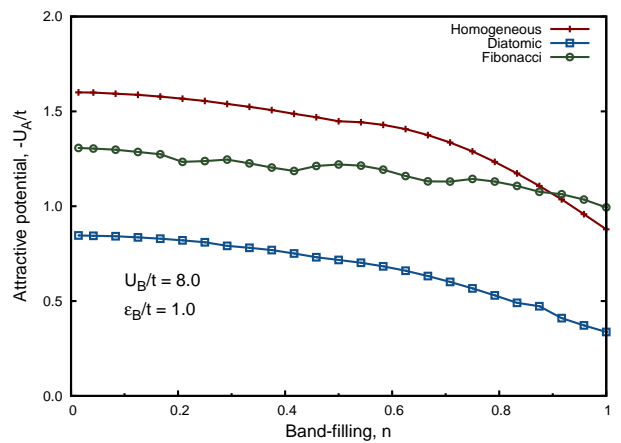}
    \caption{Minimum value of the attractive potential $U_A$
    		 for the systems to have electron pairing with
    		 a fixed repulsive potential $U_B/t=8.0$
    		 as a function of the band-filling, $n$.}
    \label{fig:min}
\end{figure}

When browsing from Figure \ref{fig:phase2} to Figure \ref{fig:phase144}, 
one observes that the electron pairing is indeed enhanced due to
quantum confinement and it also increases with the number of 
electrons in the system. 
In Figure \ref{fig:phase2} we compare our results directly to those obtained
using Bethe Ansatz (Eq. \ref{Bethe}), for the case of a binary Hubbard system 
with $\varepsilon_A-\varepsilon_B=0$ and two electrons.
In Figure \ref{fig:phase2}, it is also shown how a completely periodic
modulation of $\varepsilon$, as in the case of the diatomic chain,
enhances the most the electron pairing,  
whereas an aperiodic modulation,
such as that of the Fibonacci chain, enhances the electron pairing as
well but this is only significant when compared to the homogeneous case. 
These results are
comprehensible if we consider what was explained about the
Fibonacci chains: There are more sites with energy values
$\varepsilon_A=0$ (a total of $89$) than sites with energy values 
$\varepsilon_B=1$ (a total of $55$). This fact makes it plausible
to observe the results of the Fibonacci chain rather closer
to those of the homogeneous system than to those of the diatomic case.
The same behavior of the phase diagram was found in the systems 
at quarter-filling (see Figure \ref{fig:phase72}) where it is even
observed that the distance between the curves for both homogeneous and Fibonacci
chains decreases. This result indicates a sort of competition between
these two systems as a function of the band-filling.
An effect which close to half-filling renders interesting results.
In Figure \ref{fig:phase144} we show 
the results for the binary Hubbard chains at half-filling. While the
diatomic chain remains well below the other two cases, the
homogeneous and Fibonacci chains seem, at first glance, to behave 
the other way around, i.e., 
there are now more electrons paired for the
homogeneous than for the Fibonacci system.
We investigated further this behavior in the next way: 
After fixing a value of the repulsive on-site potential $U_B$, we 
obtained the minimum value of attractive on-site potential $U_A$ for 
which the systems have electron pairing. 
The results (see Figure \ref{fig:min}) show 
for the Fibonacci chain a uniform, 
though oscillating behavior in $U_A$. 
On the contrary, for the homogeneous case, 
and for values of the band-filling close to $n=1$, the minimum $U_A$ values
decrease faster than the Fibonacci case. 
This situation results in a critical value of the band-filling 
of $n=0.895$ after which the behavior of the homogeneous and Fibonacci 
binary Hubbard chains is inverted.  
In general, quantum confinement enhances electron pairing
in binary Hubbard chains. However, the topology of the
confinement determines strongly the electronic properties.
The behavior of the homogeneous and Fibonacci chains close
to half filling indicate that the quantum confinement effect
is overridden by the effect of the aperiodic structure, as this
is not happening in the diatomic chain. 

In conclusion, we considered a system of spin-$\frac{1}{2}$ fermions
in a one-dimensional lattice with position dependent atomic level 
$\varepsilon_i$, embedded in a binary on-site potential.
Our results illustrate how the modulation of the
atomic energy levels in a periodic and non periodic way affects the 
formation of local electron pairs
in binary Hubbard chains with attractive and repulsive 
on-site potentials. 
We found a combined effect on the electron pairing due to
the atomic level ordering and the increasing band-filling.
The number of paired-states increases with the band-filling
for the homogeneous, diatomic and Fibonacci chains. Moreover,
there is a critical value of the band-filling  
for which the effect of aperiodicity 
compared to the homogeneous case is reversed.

This work has been partially supported by Grant-57929
from CONACyT and by PAPIIT-IN108710 from UNAM. 
Y.~A. would like also to acknowledge full financial support
from grant DGAPA-UNAM.

\bibliography{literature}

\end{document}